\documentclass[aps,prl,twocolumn,groupaddress]{revtex4}

\usepackage{amsmath,amssymb,graphicx}
\usepackage{color,times}
\usepackage[latin1]{inputenc}
\usepackage{xcolor}
\usepackage{ulem}
\usepackage{afterpage}
\usepackage[T1]{fontenc}

\definecolor{darkblue}{rgb}{0, 0, 0.8}
\usepackage[colorlinks=true, breaklinks=true, linkcolor=darkblue, citecolor=darkblue, urlcolor=darkblue]{hyperref}
\newcommand{\doilink}[2]{\href{http://dx.doi.org/#1}{#2}}
\newcommand{\beq}{\begin{equation}}
\newcommand{\eeq}{\end{equation}}

\begin{document}

\title{
Optical transmission of an atomic vapor in the mesoscopic regime}

\author{T. Peyrot}
\author{Y.R.P. Sortais}
\author{J.-J. Greffet}
\author{A. Browaeys}
\affiliation{Laboratoire Charles Fabry, Institut d'Optique Graduate School, CNRS,
Universit\'e Paris-Saclay, F-91127 Palaiseau Cedex, France}

\author{A. Sargsyan}
\affiliation{Institute for Physical Research, National Academy of Sciences - Ashtarak 2, 0203, Armenia}

\author{J. Keaveney}
\author{I.G. Hughes}
\author{C.S. Adams}
\affiliation{Department of Physics, Rochester Building, Durham University,
South Road, Durham DH1 3LE, United Kingdom}

\begin{abstract}
By measuring the transmission of near-resonant light through an atomic vapor confined in a nano-cell
we demonstrate a mesoscopic optical response arising from the non-locality induced by   
the motion of atoms with a phase coherence length larger than the cell thickness.
Whereas conventional dispersion theory -- where the local atomic response is simply convolved
by the Maxwell-Boltzmann velocity distribution -- is unable to reproduce the measured spectra,
a model including a non-local, size-dependent susceptibility is found to be in excellent agreement with the measurements.
This result improves our understanding of light-matter interaction in the mesoscopic regime and has implications 
for applications where mesoscopic effects may degrade or enhance the performance
of miniaturized atomic sensors.
\end{abstract}

\maketitle

One important characteristic of mesoscopic systems is the fact that their properties are not 
ruled by local quantities. The mesoscopic regime arises when the size of the system becomes smaller 
than a distance $\xi$ characterizing the non-local response of the medium to an excitation. 
Non-locality is thus a prerequisite to the observation of mesoscopic behaviors.
For instance, the concept of local conductivity fails to describe the transport of electrons or phonons 
when the distance over which the phase of the carriers is lost
exceeds  the size of the system, as is the case in nano-wires~\cite{Agrait2003,Roukes2000}. 
In these systems the electrical potential (or the temperature) is undefined and 
one uses instead a global conductance.
Also, non-local effects are at the origin of the low temperature anomalous conductivity of a metal at frequencies ranging from
GHz to infra-red~\cite{Pippard1947}, as
the skin depth over which the field varies near a surface
is smaller than the mean-free path of the electrons in the metal~\cite{Wooten1973,Gilberd1982}.

In optics, non-locality is often observed in {\it non-linear} bulk media~\cite{Boyd,Rotschild2006}, 
in particular in the presence of long-range interactions between particles~\cite{Busche2017}. 
In contrast, manifestations of non-local optical properties in {\it linear} media are scarce. 
They have been observed for molecules near metallic surfaces~\cite{Eagen1980,Ford1984}
and the mesoscopic regime was reached with nano-particles for which the electron mean-free path is on the
order of the particle size~\cite{Kreibig1985,Voisin2001}. 
Also, the {\it selective reflection} at the interface between a glass and a bulk atomic 
vapor~\cite{Cojan1954,Burgmans1977,Briaudeau1998} was interpreted as an indirect 
evidence of non-locality originating from the motion of the atoms 
and their transient response following a collision with the glass 
surface~\cite{Schuurmans1976,Nienhuis1988,Vartanyan1994,Vartanyan1995,Ritter2018}.
Confining the vapor in nano-cells, i.e. cells with sub-wavelength thickness, the non-locality should 
give rise to a mesoscopic response, as  the system size is now on the order of the phase coherence length 
$\xi$, as explained below. These nano-cells are considered as potential atomic sensors~\cite{Knappe2005,Ritter2018} 
and ideal media to explore atom-light~\cite{Keaveney2012} and atom-surface interactions~\cite{Sargsyan2017}. 
It is therefore important to understand how the mesoscopic response may affect the precision of these sensors. 
So far, the interplay between non-locality and system-finite size have been very little studied in nano-cells~\cite{Dutier2003b}
with no comparison between experiment and theory. In particular, the question remains whether 
the concept of susceptibility holds in this mesoscopic system.

Here, we systematically study the mesoscopic optical response of 
a hot vapor of cesium confined in a nano-cell.
We measure the transmission spectra for various cell thicknesses and 
observe that they cannot be reproduced by a model assuming a local susceptibility. 
We develop a theoretical model where we calculate explicitly the mesoscopic optical response of the 
vapor accounting for the non-locality arising from the motion of the atoms and for the breaking 
of translational invariance due to the presence of interfaces. In particular, our model defines clearly 
the parameter regime (velocity, density, and size of the system) where non-locality 
dominates in atomic vapors, identifies the role of the system finite-size and reproduces 
remarkably well the observed spectra.

\begin{figure}
\includegraphics[width=\columnwidth]{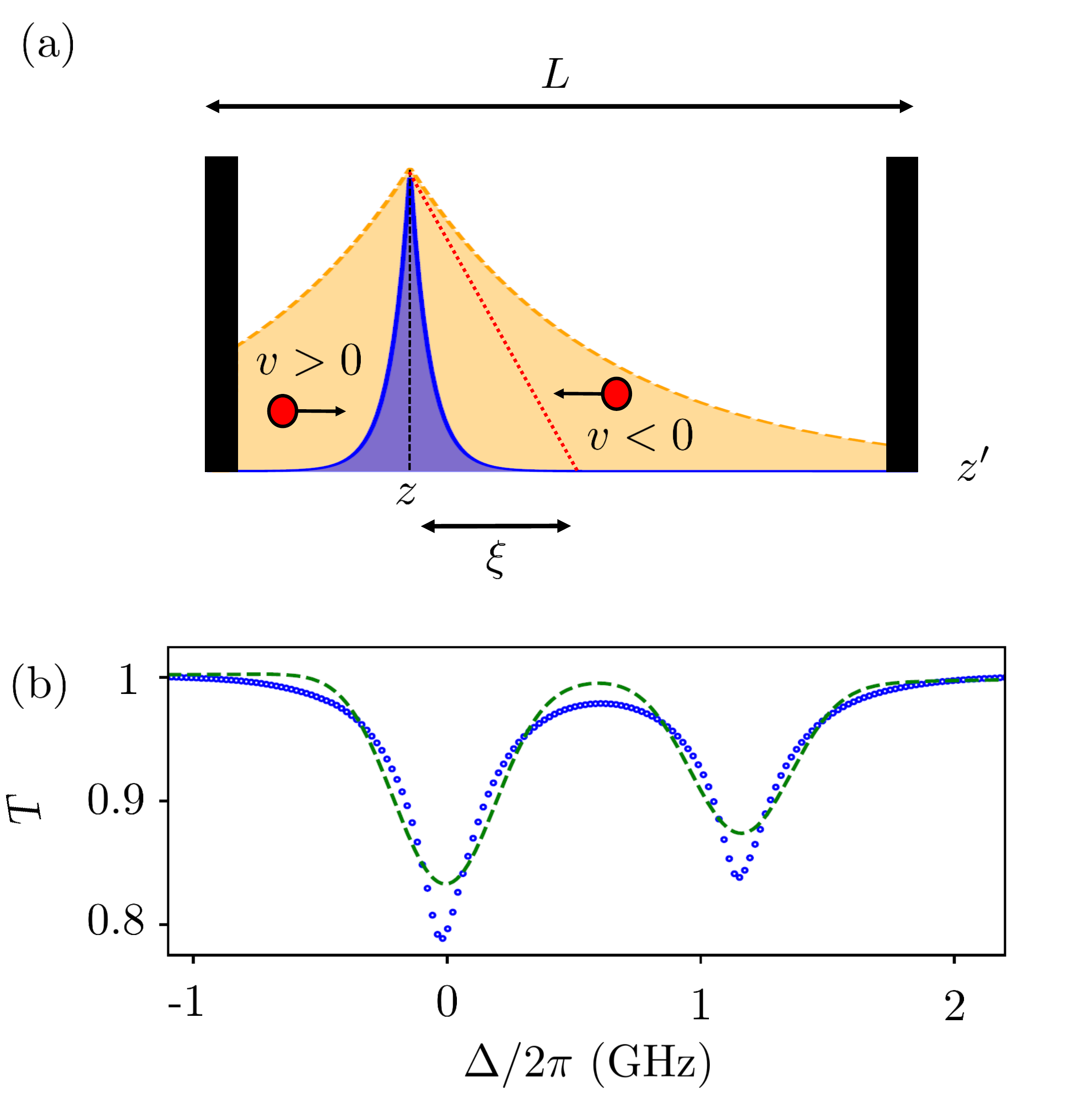}
\caption{(a) Illustration of the non-local response in presence of an interface in a slab of thickness $L$.
Orange fill: non local response $\chi$ for $\xi\sim L$. Blue fill : local response $\chi$ for $\xi \ll L$.
(b) Blue line: experimental transmission spectrum for $L= 420$\,nm and 
$\Theta\approx 170^{\circ}$C as a function of the
laser detuning $\Delta$ with respect to the transition $F=4 $ to $ F'=3$.
The data are binned 10 times by steps of $2$\,MHz. The lines correspond respectively to the
Cs D1 hyperfine transitions
$F=4$ to $F'=3$ (left), and $F=4$ to $F'=4$ (right).
Dashed green line: fit by the first local model.}
\label{Fig1}
\end{figure}

In any {\it homogeneous} medium, the relation between the polarization vector and the electric field at 
a frequency $\omega$ is given by (one-dimensional model)~\cite{LandauE&M}:
\beq\label{Eq:NonLocalPola}
P(z,\omega)=\int_{-\infty}^{+\infty}\epsilon_0\chi(z-z',\omega)E(z')dz'\ ,
\eeq
where the susceptibility $\chi(z-z',\omega)$ describes the spatial response of the medium 
and typically decays over a distance $\xi$, the so-called range of non-locality.
In an atomic vapor, the non-locality comes from the motion of the atoms and 
$\xi$ is equal to their phase coherence length, 
i.e. the distance travelled by the atoms before the phase of the light excitation imprinted on 
them is lost (due to collisions with other particles or to radiative decay): $\xi=v/\Gamma_{\rm t}$ 
with $v$ the atom velocity and $\Gamma_{\rm t}$ the total homogeneous 
linewidth~\cite{Schuurmans1976,Nienhuis1988,Vartanyan1995}. Typically, in a room temperature vapor of
alkali, $\xi\approx 3$\,$\mu$m. In a nano-cell, the presence of the 
walls separated by $L$ breaks the translational invariance (see Fig.\ref{Fig1}(a)) and the mesoscopic regime is
achieved as soon as $\xi \gtrsim L$. In this regime, the non-local relation between 
$P$ and $E$ (Eq.\,\eqref{Eq:NonLocalPola}) 
also depends on $L$ and the concept of size-independent susceptibility collapses, as is also 
the case in nano-photonic devices~\cite{Eagen1980,Ford1984,Cocoletzi2005,Churchill2016,Tserkezis2017}.

To observe the mesoscopic optical response resulting from non-locality, we confine a Cs vapor in a wedged sapphire 
nano-cell of refractive index $n_{\rm s}=1.76$, the thickness of which varies from $30$\,nm to 
$2\,\rm \mu$m~\cite{Sargsyan2001,Peyrot2018}. 
The cell is mounted in a home-made oven that allows  differential heating between the reservoir and the windows.
The reservoir temperature $\Theta$ is monitored by a thermocouple and is 
related to the atomic density $N$ in the cell via the vapor pressure. 
An external-cavity diode laser is scanned at $10$\,Hz around the Cs D1 line at $\lambda = 894$\,nm 
(natural linewidth $\Gamma=2\pi\times 4.6$\,MHz) and we use a $7$\,cm spectroscopic cell as a 
reference for frequency calibration. 
The $700$\,nW laser beam is focused with a waist of $\sim40\,{\rm \mu}$m and scanned 
along the wedge to explore various thicknesses $L$ of the atomic slab. We use
the back reflections on the nano-cell to determine $L$ using an interferometric 
method~\cite{Bouchiat2000}. Finally, the transmitted light is collected on a photodiode.

When the temperature of the vapor increases, so do the density and the homogeneous 
linewidth due to collisional dipole-dipole interactions. For $\Theta\geq 250^\circ$C, 
we observe linewidths as large as $\Gamma_{\rm t}= 2\pi\times 1$\,GHz 
leading to 
$\xi<L$, thus restoring locality~\cite{footnote_Peyrot2018}. 
To avoid this situation, we set the temperature of the vapor to a  lower
value ($\Theta\approx170^{\circ}$C) to keep the expected homogeneous linewidth 
$\Gamma_{\rm t}\approx2\pi\times 60$\,MHz~\cite{Weller2011} such that 
$\xi>5L$. Operating at a lower temperature would  
make the mesoscopic response stronger, at the expense of a much lower absorption, hence reducing the signal-to-noise ratio. 
The choice of the temperature thus results from a compromise. We present in Fig.\,\ref{Fig1}(c) 
an example of a transmission spectrum, normalized to the value of the signal far from the 
atomic resonances, for a thickness of the slab $L=420$\,nm. The lineshape appears more 
complicated than a sum of Gaussian or Lorentzian functions.

In an attempt to model the transmitted spectra, we first use the dispersion theory relying on a description 
of the vapor (density $N$) by a {\it local} susceptibility $\chi$~\cite{f2f}. 
Specifically, we calculate $\chi$ by summing the contributions of all Doppler-broadened 
hyperfine transitions of the Cs D1 line at frequencies $\omega_{FF'}$~\cite{ElecSus}, 
assuming a normalized Maxwell-Boltzmann velocity distribution $M_v(v)$ along the laser 
direction of propagation ($d$ is the dipole moment of the strongest transition, 
$C_{FF'}$ the Clebsch-Gordan coefficients):
\beq\label{Eq:susceptibilityELECSUS}
\chi (\omega)=
{Nd^2\over \hbar \epsilon _0}
\sum_{F,F'}C_{FF'}^2\int_{-\infty}^{\infty} \frac{iM_v(v)}{\Gamma_{\rm t}-2i(\Delta_{FF'}-k_{\rm l}v)}dv\ .
\eeq
Here, $\Delta_{FF'}=\omega- \omega_{FF'} -\Delta_{\rm p}$ and $\Gamma_{\rm t}=\Gamma+\Gamma_{\rm p}$, 
where we have introduced $\Delta_{\rm p}$ and $\Gamma_{\rm p}$ a shift and a broadening characterizing the medium, 
which originates from the collisional dipole-dipole interactions and the  atom-surface interactions~\cite{Peyrot2018}. 
The refractive index is then $n(\omega)=\sqrt{1+\chi(\omega)}$. We account for the multiple reflections 
inside the cavity formed by the two sapphire plates (index $n_{\rm s}$) surrounding 
the vapor using  the transmission function:
\beq\label{Eq:FPtrans}
t(\omega)={4 n_{\rm s} n\exp[i(n-n_{\rm s})k_{\rm l} L]\over (n_{\rm s}+n)^2-(n_{\rm s}-n)^2 \exp[2in k_{\rm l} L]}\ .
\eeq
Finally, we calculate the normalized transmission $T(\omega)=|t[n(\omega)]/t[n=1]|^2$.
The result of this  first model is shown in Fig.\,\ref{Fig1}(c), for which we have adjusted the
values of $N$, $\Delta_{\rm p}$ and $\Gamma_{\rm p}$ to best fit the data.
Strikingly, it does not agree with the data:
the experimental linewidth appears narrower than the calculated Doppler broadened width.
This is a signature of the coherent Dicke narrowing
already observed by many authors~\cite{Dicke1953,Dicke1955,Dutier2003b,Sargsyan2016}.
In nano-cells, this emphasizes the failure of the conventional dispersion theory, which assumes a local
susceptibility of the atomic gas and a Maxwell-Boltzmann velocity distribution.

\begin{figure}
\includegraphics[width=\columnwidth]{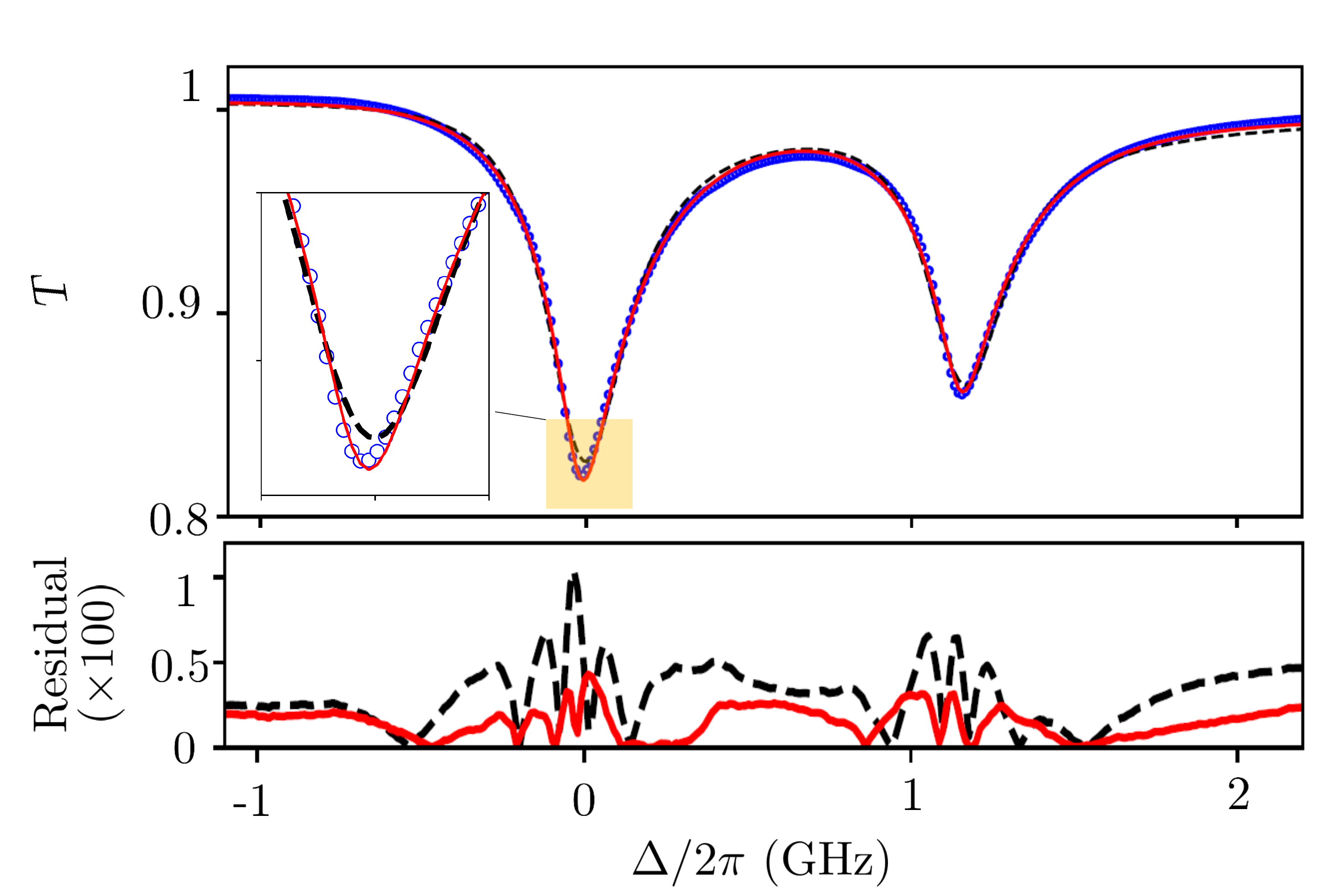}
\caption{Top panel.
Blue dots:  experimental transmission spectrum for
$\Theta=170^{\circ}$C and $L=360$\,nm binned as in Fig.\,\ref{Fig1}(c).
Black dot-dashed line: fit by the second model.
Red dashed line: fit by the third model (see text).
Inset: zoom near resonance. Low panel: absolute value of the residuals for both fits.
}
\label{Fig2}
\end{figure}

In a second model, we introduce the effect of the cell walls in the simplest possible way:
we assume that mainly the atoms flying parallel to the walls contribute to the signal, 
all the others colliding too rapidly with the walls to participate. 
We therefore take for the velocity distribution $M_v(v)=\delta(v)$ in 
Eq.\,(\ref{Eq:susceptibilityELECSUS})~\cite{footnote_bimodal}, to account phenomenologically 
for the velocity selection. We fit the data letting as before $N$, $\Delta_{\rm P}$ and $\Gamma_{\rm P}$ 
free to evolve. The result, shown in Fig.\,\ref{Fig2} for $L=360$\,nm, is in much better agreement with the data. 
Nonetheless, the residuals reveal that the model fails to reproduce the narrow feature 
near resonance, characteristic of the contribution from the slow atoms~\cite{Briaudeau1998}.

Finally, inspired by previous works~\cite{Vartanyan1995,Nienhuis1997,Dutier2003a}, we derive a third, 
intrinsically non-local model that accounts both for the explicit $k$-dependence 
of the susceptibility and the collisions of the atoms with the surfaces of the nano-cell. 
To do so we first calculate the response function of the atomic medium assuming it is homogeneous
and then we account for the influence of the surfaces~\cite{Ford1984}. 
The susceptibility in the $(k,\omega)$ space of an homogeneous gas of atoms with a velocity $v$ is given, 
for a specific transition, by~\cite{SM}:
\beq \label{Susceptibility1}
\chi_{FF'}(k,\omega,v)=i{d^2C_{FF'}^2\over \hbar\epsilon_0}{N M_v(v)\over\Gamma_{\rm t}- 2i(\Delta_{FF'}-kv)}\ .
\eeq
The $k$-dependence resulting from the Doppler effect 
is at the origin of the non-locality and leads to spatial dispersion~\cite{LandauE&M}. Note that 
this non-locality is not specific to nano-cells, but appears in any atomic vapor.
An inverse Fourier transform yields:
\begin{equation} \label{chi1}
\chi_{FF'}(z-z',\omega,v)= iNM_v(v)\frac{d^2C_{FF'}^2}{\hbar\epsilon_0 |v|}~
e^{(-{\Gamma_{\rm t}\over 2}+i\Delta_{FF'}){z-z'\over v}}
\end{equation}
for $(z-z')/v>0$ and $\chi_{FF'}(z-z',\omega,v)=0$ for $(z-z')/v<0$, as required by causality. 
We recover the above-mentioned decay length $\xi=|v|/\Gamma_{\rm t}$. 
As for the influence of the surfaces, we assume quenching collisions with the cell walls~\cite{Schuurmans1976}, 
i.e. the phases of the atomic coherences are reset upon collisions. 
Velocity classes $\pm v$ become independent and  
the presence of the walls therefore breaks the translational invariance in the medium~\cite{footnote_elastic_collisions}. 
We express this fact by multiplying $\chi_{FF'}(z-z',\omega,v)$ by a top-hat function 
($\Pi_{0}^{L}(z')=1$ for $0<z'<L$ and is null elsewhere). 
When $\xi\gtrsim L$, the non-local response of the medium depends on the size $L$ of the entire system, 
and is not characteristic of the medium only (see Fig.\,\ref{Fig1}(b)). 
Finally, the response of the system is obtained by summing over all the atomic transitions: 
$\chi_{_L}(z,z',\omega,v)=\sum_{F,F'}\Pi_{0}^{L}(z')\times\chi_{FF'}(z-z',\omega,v)$.

To calculate the field transmitted through the cell filled with the vapor, 
we also consider the multiple reflections inside the cavity formed by the sapphire windows.
The transmitted field $E_{\rm t}$ is the superposition of the field transmitted by the empty cavity $E_{\rm t0}$, 
and of the fields $E_{\rm t+}$ and $E_{\rm t-}$ initially scattered by the atoms in the forward 
and backward directions and that have undergone multiple reflections before being transmitted.
The field transmitted by the empty cavity is $E_{t0}=t_1t_2/(1-r_2^2e^{2i k_{\rm l} L})E_0 e^{i k_{\rm l} z}$ with
$t_1=2n_{\rm s}/(1+n_{\rm s})$, $t_2=2/(1+n_{\rm s})$, $r_2=(1-n_{\rm s})/(1+n_{\rm s})$
and $E_0 e^{i k_{\rm l} z}$ the incident field.
The fields $E_{t+}$ and $E_{t-}$ are related to the polarization vector $P(z,\omega)$
inside the medium by~\cite{SM}:
\begin{align}
E_{t\pm}(z)={t_2\over1-r_2^2e^{2i k_{\rm l} L}}\frac{i k_{\rm l} }{2\epsilon_0}\int_{0}^{L}dz' P(z',\omega)e^{i k_{\rm l} (z\mp z')}\ ,
\end{align}
where the polarization inside the medium is linked to the {\it total} cavity field $E(z)$ 
by the expression generalizing Eq.\,\eqref{Eq:NonLocalPola}:
\beq
P(z',\omega)=\int_{-\infty}^{\infty}dz''
\int_{-\infty}^{\infty}dv\  \epsilon_0 \chi_{_L}(z',z'',\omega,v)E(z'')\ .
\eeq
The integrals can be calculated~\cite{SM}, assuming the atomic medium to be dilute 
and thin enough so that the cavity field is approximately the one inside the empty cavity
(Born approximation~\cite{Carminati1995}) : 
$E(z'')\approx t_1/ (1-r_2^2e^{2i k_{\rm l} L})E_0[e^{i k_{\rm l} z''}+r_2e^{i k_{\rm l} (2L-z'')}]$. 
Under this assumption, and taking a
Maxwell velocity distribution, we compute them numerically to extract the normalized 
transmission $T(\omega)=|E_{\rm t}/E_{\rm t0}|^2$. After abandoning the description of the system by a size-independent
response function, the transmission is now the global observable characterizing the optical response in our mesoscopic regime. 
Our approach starting from the non-local response function agrees with the formulae
obtained in Ref.~\cite{Dutier2003a} under the same assumption, i.e. for low absorption.

\begin{figure}
\includegraphics[width=\columnwidth]{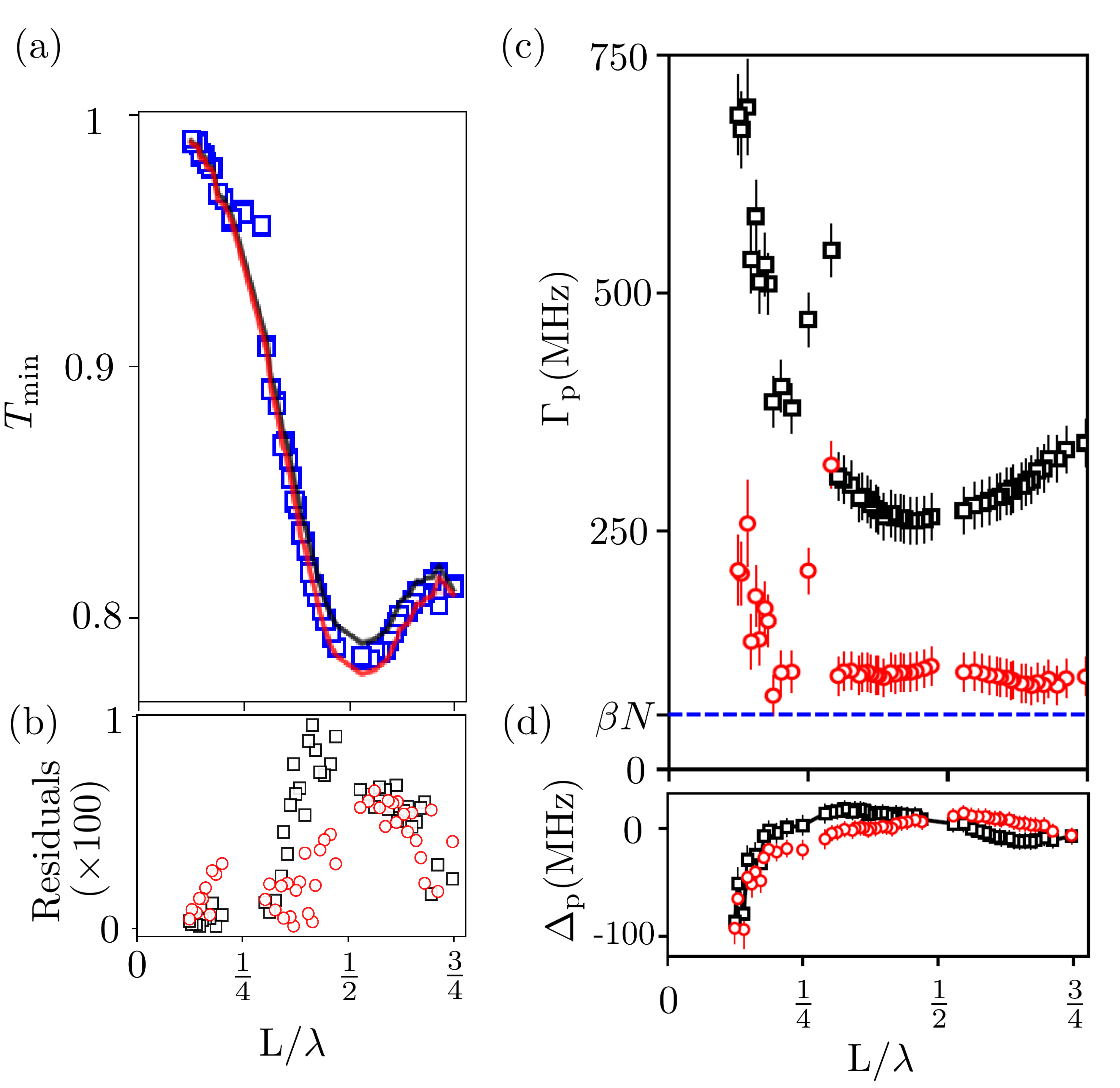}
\caption{(a) Minimum of transmission $T_{\rm{min}}$ for the
transition from $F=4$ to $F'=3$ against the cell thickness $L$.
Blue squares: experimental data.
Black and red lines: values deduced from the fit of the spectra using the second and third models, respectively.
Error bars are smaller than markers.
(b) Absolute value of the residuals for the second (black) and third (red) models.
(c) (resp. d): broadening parameter $\Gamma_{\rm p}$
(resp. shift parameter $\Delta_{\rm p}$) obtained from the fit of
the spectra using the second model (black  squares) and third model (red  circles).
Error bars are the quadratic sum of statistic and fit errors.
Blue dotted line: collisional broadening prediction~\cite{Weller2011}. }
\label{Fig3}
\end{figure}

The  fit of the data by the third model is presented in Fig.\,\ref{Fig2} for the best 
found parameters $N$, $\Delta_{\rm p}$ and $\Gamma_{\rm p}$.
The agreement is excellent. In particular, the narrow feature near resonance is reproduced accurately: 
despite the fact that we keep the full Maxwell velocity distribution, the velocity selection, 
imposed in the second phenomenological model and at the origin of the narrowing, 
is an automatic consequence of the third, non-local model.
To further test the two last models, we also plot in Figs.\,\ref{Fig3}(a,b) the value $T_{\rm{min}}$ of the 
minimum of the transmission for the hyperfine transition from $F=4$ to $F'=3$
as a function of the cell thickness. We observe that both models are in good agreement with the data 
although the third model fits better around
$L\approx\lambda/2$~\cite{Footnote_lambda4}. Also, $T_{\rm min}(L)$ does not 
decay exponentially as the Beer-Lambert law would predict~\cite{f2f}. This is expected for two reasons.
Firstly, the atoms being in a cavity, the transmitted field amplitude is not given by the Beer-Lambert law 
but by Eq.\,\eqref{Eq:FPtrans}: a $\lambda/2$-periodic oscillation, originating from the multiple reflections 
in the cavity, modulates the exponential decay. Secondly, even without the cavity, the field inside the vapor 
cannot be exponential due to the non-local character of the medium~\cite{Schuurmans1976}, 
which leads to a $\lambda$-periodic oscillation~\cite{SM}.

Even though the residuals in Fig.\,\ref{Fig2} could discriminate between the second phenomenological 
and third non-local models~\cite{Hughes}, the values of
$\Delta_{\rm p}$ and $\Gamma_{\rm p}$ returned by the fit indicate 
clearly that only the third model is correct, as we now discuss. 
Both parameters characterize the bulk properties of the vapor and the 
van der Waals interactions between the atoms and the surfaces. 
They depend {\it a priori} on the density $N$ (constant at a given temperature of the vapor) and the cell thickness $L$.
The $L$-dependence comes only from the atom-surface interaction, as for small $L$ 
the fraction of atoms close to the surface is larger than for large $L$.
For Cs, the theoretical atom-sapphire interaction coefficient $C_3$
is around a few kHz.$\mu$m$^3$~\cite{Ducloy2005,Whittaker2014}:
in the range  $\lambda/4\leq L \leq\lambda$ the influence of the surface
on $\Gamma_{\rm p}$ and $\Delta_{\rm p}$ is therefore expected to be smaller than 10\,MHz and thus  negligible.
Importantly, the cavity effects are already taken
into account in both models through the multiple reflections
and therefore should not contribute to $\Gamma_{\rm p}$ and $\Delta_{\rm p}$~\cite{Footnote_CLS}.
For a fitting model to make sense, it should therefore return values of $\Gamma_{\rm p}$
and $\Delta_{\rm p}$ independent of $L$.
Figure~\ref{Fig3}(c) shows the values of $\Gamma_{\rm p}$ returned by the fit
for the two last models as a function of $L$. 
Only the third model is able to return a value independent of $L$ for $L\ge 200$\,nm.
At smaller distances, $\Gamma_{\rm p}$ increases due to the atom-surface interaction~\cite{SM}.
Furthermore, for $L\ge 200$\,nm, $\Gamma_{\rm p}$
is in reasonable agreement with the expected broadening $\beta N$ due to collisional dipole-dipole
interactions at  the density corresponding to
$\Theta\approx170^{\circ}$C~\cite{Weller2011}.
The second, phenomenological model, by contrast, yields a strong
dependence of $\Gamma_{\rm p}$ with $L$, which is not acceptable based on the arguments presented
above. As a consequence the only model, which features both a good agreement with the data and
a consistent interpretation of its fitting parameters, is the third one.
Figure~\ref{Fig3}(d) presents the fitted $\Delta_{\rm p}$ against $L$:
the difference between the two models is less striking. Both feature the influence of the
attractive atom-surface  interaction at small thickness.

The situation studied in this work, of an atomic vapor where the phase coherence length 
exceeds the dimension of the system, is widely met in miniaturized atomic sensors. 
We have shown that the propagation of light
through nano-cells cannot be described by any local property, and even the concept of 
non-local system-size-independent susceptibility collapses. The optical response of mesoscopic 
systems is now understood globally using a transmission factor. 
This situation is reminiscent of the electrical conduction in the mesoscopic regime 
where the concept of local conductivity is no longer valid and a global conductance has to be introduced. 
Our model makes explicit the role of non-locality and its dependence with the system size, 
and agrees with experimental data for extinctions as large as $20\%$. 
To our knowledge the agreement presented here between theory and experiment is unprecedented 
in both atomic hot and cold dense atomic vapors altogether~\cite{Jennewein2018}. 
Importantly, it allows the extraction of meaningful quantities such as energy shift and linewidth, 
hence providing a theoretical framework for characterizing future atomic sensors.

\begin{acknowledgments}
We thank D.~Sarkisyan, G.~Dutier, A.~Laliotis and D.~Bloch for discussions. 
T.~Peyrot is supported by the DGA-DSTL fellowship 2015600028. 
We also acknowledge financial support from CNRS, EPSRC (grant EP/R002061/1) and Durham University. 
The data presented in this paper  will be available later on.
\end{acknowledgments}


\begin{thebibliography}{50}
\bibitem{footnote_elastic_collisions}
This is in contrast to the case of elastic collisions at the wall~\cite{SM}.
\end{thebibliography}
\end{document}


\title{Supplemental Material: Optical transmission of an atomic vapor in the mesoscopic regime}

\author{T. Peyrot}
\author{Y.R.P. Sortais}
\author{J.-J. Greffet}
\author{A. Browaeys}
\affiliation{Laboratoire Charles Fabry, Institut d'Optique Graduate School, CNRS,
Universit\'e Paris-Saclay, F-91127 Palaiseau Cedex, France}

\author{A. Sargsyan}
\affiliation{Institute for Physical Research, National Academy of Sciences - Ashtarak 2, 0203, Armenia}

\author{J. Keaveney}
\author{I.G. Hughes}
\author{C.S. Adams}
\affiliation{Department of Physics, Rochester Building, Durham University, South Road, Durham DH1 3LE, United Kingdom}


\maketitle

Here, we first derive the equations of the third (non-local) model of the main text,
starting with the expression of the non-local susceptibility (Eq.\,(5)).
Then, we derive the expression of the optical field transmitted through a thin layer of atomic gas,
first in vacuum, and then accounting for the presence of glass interfaces.
In Section~\ref{Sec:bimodal}, we show that using a bimodal velocity distribution in the second
model of the main text cannot reproduce consistently the data either.
In Section~\ref{Sec:Beer} we give more details about the deviation of the transmission
coefficient from the Beer-Lambert law.
Finally, we analyze in more details the thickness dependence of the parameter
$\Gamma_{\rm p}$ extracted from the fit of the data by the local model.

\section{Derivation of the non-local model}\label{Sec:NLmodel}

In this first Section, we derive the expression of the transmission through the atomic slab, including the cavity effect
from the sapphire plates of the nanocell, making explicit the origin of the non-locality.

\subsection{The non-local susceptibility}\label{SubSec:NLchi}

We start by deriving the expression of the non-local
susceptibility $\chi (z,z',\omega,v)$ of the ensemble of atoms
for a velocity class $v$ and at frequency $\omega$. As explained in the main text,
the presence of the cell walls makes it a challenging task in general.
To simplify the situation, we will use the following procedure:
\begin{enumerate}

\item We will treat the vapor as if it was homogeneous (i.e. in the absence of confining walls).
In this case, the susceptibility depends only on the difference: $\chi (z,z',\omega,v)=\chi (z-z',\omega,v)$.

\item We will treat the effect of the surfaces separately. This will be done by assuming quenching collisions at the cell walls,
i.e. the atomic coherence will be lost during these collisions~\cite{Schuurmans1976} and reset to zero.
In this way, there will not exist any relation between the coherence of atoms moving at $+v$ or $-v$,
contrarily to what would happen if the collisions with the walls were elastic.

\end{enumerate}

To derive the susceptibility for an homogeneous vapour, we first consider a given atomic transition
between ground and excited states $F$ and $F'$
(frequency $\omega_{FF'}$, and total homogeneous linewidth $\Gamma_{\rm t}$) described by a coherence $\rho_{21}$
and a dipole matrix element $d_{FF'}=C_{FF'}\,d$ ($C_{FF'}$ is the Clebsch Gordan coefficient
normalized to the strongest transition with dipole moment $d$).
We write the field and polarization in the vapor in the form:
$E(z,t)=\frac{1}{2}\left(E(z)e^{-i\omega t}+E^*(z)e^{i\omega t}\right)$ and $P(z,t)=\frac{1}{2}\left(P(z)e^{-i\omega t}+P^*(z)e^{i\omega t}\right)$.
We introduce the coherence field $\rho_{21}(z,t,v)$ at position $z$ in the slab for the velocity class $v$.
In the low intensity limit (i.e. neglecting populations in the excited state), the Maxwell-Bloch equation for the coherence is~\cite{Aspect2010}:
\beq \label{Maxwell-Bloch1}
\frac{{\rm d} \rho_{21}(z,t,v)}{{\rm d} t} =-i\omega_{FF'} \rho_{21}(z,t,v)+iC_{FF'}\frac{dE(z,t)}{\hbar}-\frac{\Gamma_{\rm t}}{2} \rho_{21}(z,t,v)\ .
\eeq

Using the convective derivative ${\rm d}/  {\rm d} t=\partial/ \partial t+v\partial /\partial z$, we get:
\beq \label{Maxwell-Bloch2}
\frac{\partial \rho_{21}(z,t,v)}{\partial t}+v\frac{\partial \rho_{21}(z,t,v)}{\partial z} =
-i\omega_{FF'} \rho_{21}(z,t,v)+iC_{FF'}\frac{dE(z,t)}{\hbar}-\frac{\Gamma_{\rm t}}{2} \rho_{21}(z,t,v)\ .
\eeq
Taking the Fourier transform of this equation with respect to $t$ and $z$, in the quasi-resonant approximation, we obtain:
\beq \label{Coherences}
\rho_{21}(k,\omega,v)=iC_{FF'}{dE(k,\omega)\over2\hbar}{1\over\Gamma_{\rm t}/2-i(\Delta_{FF'}-kv)}\ ,
\eeq
with $\Delta_{ FF'}=\omega-\omega_{ FF'}$. In the $(k,\omega)$ space and for an {\it homogeneous medium},
the relation between the polarization and the field is:
$P(k,\omega,v)=\epsilon_0\chi(k,\omega,v)E(k,\omega)$.
Furthermore, in the dilute approximation, $P(k,\omega,v)= 2N\,d\,C_{FF'}M_v(v)\rho_{21}(k,\omega,v)$,
with $N$ the density of the vapor and $M_v(v)$ the Maxwell Boltzman velocity distribution.
Consequently the susceptibility  of the vapour in the $(k,\omega)$ space is:
\beq \label{Susceptibility1}
\chi(k,\omega,v)=iNM_v(v){C_{FF'}^2d^2\over \hbar\epsilon_0}{1\over\Gamma_{\rm t}/2-i(\Delta_{FF'}-kv)}\ .
\eeq
This susceptibility depends explicitly on $k$ through the Doppler shift,
and this is the reason for the non-local response of the medium (i.e. the spatial dispersion).
We note that this approach
using a convective derivative is similar to the one used to derive the permittivity of a metal
in  the anomalous skin-depth situation~\cite{Gilberd1982}.
The inverse  Fourier transform in $k$ yields the susceptibility:
\beq \label{Susceptibility2}
\chi(z-z',\omega,v)= iNM_v(v){C_{FF'}^2d^2\over \hbar\epsilon_0}{1\over 2\pi}
\int_{-\infty}^{\infty}{e^{ik(z-z')}\over \Gamma_{\rm t}/2-i(\Delta_{FF'}-kv)}\, dk\ .
\eeq
To calculate the integral, we integrate in the $k$-complex plane.
When applying the residue theorem,
care must be taken, as the position of the pole $k_0=(\Delta_{FF'}+i\Gamma_{\rm t}/2)/v$
in the plane depends on the sign of the velocity.
We obtain the non-local response function presented in the main text [Eq.(5)]:
\begin{align}
&\chi(z-z', \omega, v)= 0 \label{chi1}\ \ \ \ {\rm when} \ \ \ \ {z-z'\over v} <0\ , \\
&\chi(z-z',\omega, v)=iNM_v(v)\frac{C_{FF'}^2d^2}{\hbar\epsilon_0 |v|}\exp\left[\left(-{\Gamma_{\rm t}\over 2}+i\Delta_{FF'}\right){z-z'\over v}\right]
\ \ \ \ {\rm when} \ \ \ \ {z-z'\over v} >0\ . \label{chi2}
\end{align}
The expression~(\ref{chi2}) shows that the distance $\xi$ over which the polarization
depends on the field (i.e. the range of non-locality) is $\xi = |v|/\Gamma_{\rm t}$ ,
thus confirming the qualitative discussion in the main text. Furthermore the fact that
$\chi(z-z',\omega,v)$ depends on the sign of $(z-z')/v$ indicates that the optical response
at position $z$ only depends on atoms located at position $z'<z$ moving with a positive velocity
and on atoms at $z'>z$ having a negative velocity.

\subsection{Transmitted field for a dilute slab placed in vacuum}\label{Sec:slab_vacuum}

We first consider an atomic slab of thickness $L$ placed in vacuum, i.e. we do not consider the sapphire windows of the nano-cell. T
he slab is excited by a plane wave (frequency $\omega$) with complex amplitude $E_0\exp[ik_{\rm l}z]$
where $k_{\rm l}=\omega / c$. Using the superposition principle,
the amplitude of the field transmitted after the slab is~\cite{Milonni1996,Feynman}:
\beq\label{Eq:Superposition_principle}
E_{\rm t}(z>L) = E_{0} e^{i k_{\rm l} z} + {ik_{\rm l}\over 2\epsilon_{0}}\int_0^L dz' P(z',\omega) e^{ik_{\rm l}(z-z')}\, \ .
\eeq
In a non-local linear  medium consisting of atoms moving with velocity $v$ (distribution $M_v(v)$) as considered here,
the relation between the polarization vector and the field is:
\beq\label{Eq:NonLocalPola}
P(z,\omega)=\int_{-\infty}^{\infty}dz'\int_{-\infty}^{\infty}dv\, \epsilon_0\chi(z,z',\omega,v)\,E(z')\ ,
\eeq
leading to the following expression for the transmitted field:
\beq\label{trans1}
E_{\rm t}=E_0e^{ik_{\rm l}z}+\frac{ik_{\rm l}}{2\epsilon_0}\int_{0}^{L}dz'e^{ik_{\rm l}(z-z')}
\left[ \int_{-\infty}^{\infty} dz'' \epsilon_0 \left(\int_{-\infty}^{\infty}dv \,\chi(z',z'',\omega,v) \right)E(z'')\right]\ .
\eeq
The hypothesis of quenching collisions at the edge of the slab (second hypothesis
at the beginning of Sec.\,\ref{SubSec:NLchi})  leads to a breaking of the translational invariance
by the interfaces, a fact that we express
by multiplying the susceptibility of \eqref{chi2} by a top-hat function ($\Pi_{0}^{L}(z')=1$ for $0<z'<L$ and is null elsewhere):
$\chi_{_L}(z',z'',\omega,v)=\chi(z'-z'',\omega,v)\Pi_{0}^{L}(z'')$.
Using the expression of the non-local susceptibility \eqref{chi2},
the second term of \eqref{trans1} is a sum of two integrals:
\begin{align}
I_0&=\frac{ik_{\rm l}}{2\epsilon_0}\int_{0}^{L}dz'e^{ik_{\rm l}(z-z')}
\left[\int_{0}^{z'}dz'' \epsilon_0 \left( \int_{0}^{\infty} dv\,
\frac{iNC_{FF'}^2d^2}{\hbar\epsilon_0 v}e^{{z'-z''\over v}(-{\Gamma_{\rm t}\over2}+i\Delta_{FF'})} M_v(v) \right) E(z'')\right]  \label{I1}\\
I_0'&=\frac{ik_{\rm l}}{2\epsilon_0}\int_{0}^{L}dz'e^{ik_{\rm l}(z-z')}
\left[\int_{z'}^{L}dz''\epsilon_0 \left( \int_{-\infty}^{0}dv\,
\frac{-iNC_{FF'}^2d^2}{\hbar\epsilon_0 v}e^{{z'-z''\over v}(-{\Gamma_{\rm t}\over2}+i\Delta_{FF'})}M_v(v) \right)E(z'') \right] \ .\label{I2}
\end{align}

The separation of the integral in Eq.\,\eqref{trans1}
into two integrals, one involving negative velocities, and the other positive
ones, using Eqs.\,\eqref{chi1} and \eqref{chi2} for $\chi(z'-z'',\omega,v)$ also comes from the assumptions
that the atoms lose their coherence
at the boundaries of the slab. Otherwise, an atom with a velocity $-v$ would bounce off the surface,
switching its velocity to $+v$, and would contribute to the polarization even for $z'<z$.

In order to proceed further, we need the expression of the field inside the vapor.
As we consider a dilute atomic medium placed in vacuum, we neglect the reflection at the
boundaries of the slab and take $E(z'')\approx E_0\exp[ik_{\rm l}z'']$. The integrals $I_0$ and $I_0'$  become respectively:
\begin{align}
I_0&=-{NC_{FF'}^2d^2\over2\hbar\epsilon_0 }E_0\exp[ik_{\rm l}z]\int_{0}^{\infty} dv M_v(v) \left( \frac{k_{\rm l}L}{\Lambda}+
\frac{k_{\rm l}v\, e^{-{\Lambda L\over v}}}{\Lambda^2}-\frac{k_{\rm l}v}{\Lambda^2}\right)\ , \label{I1fin} \\
I_0'&=-{NC_{FF'}^2d^2\over2\hbar\epsilon_0 }E_0\exp[ik_{\rm l}z]\int_{-\infty}^{0} dv M_v(v) \left( \frac{k_{\rm l}L}{\Lambda}-
\frac{k_{\rm l}v \, e^{{\Lambda L\over v}}}{\Lambda^2}+\frac{k_{\rm l}v}{\Lambda^2}\right) \ , \label{I2fin}
\end{align}
where we have introduced, as in Ref.~\cite{Briaudeau1998}, $\Lambda=\Gamma_{\rm t}/2-i(\Delta_{FF'}-k_{\rm l}v)$.
Combining Eqs.\eqref{trans1}, \eqref{I1fin} and \eqref{I2fin}, the final expression of the  field transmitted after the slab is:
\beq\label{transFin1}
E_{\rm t}=E_0e^{ik_{\rm l}z} \left[1-\frac{NC_{FF'}^2d^2}{2\hbar\epsilon_0 }\int_{-\infty}^{\infty} dv M_v(v)
\left( \frac{k_{\rm l}L}{\Lambda}-\frac{k_{\rm l}|v|}{\Lambda^2}+\frac{k_{\rm l}|v| e^{-{\Lambda L \over |v|}}}{\Lambda^2}\right) \right]\ ,
\eeq

Equation\,\eqref{transFin1} is equivalent to the one derived by several authors~\cite{Vartanyan1995,Briaudeau1998}
starting from a different point of view, which made the non-locality less explicit than the above derivation.
We now discuss the contributions of the three integrals (labeled $J_1, J_2$ and $J_3$ respectively)
in Eq.\,\eqref{transFin1} (see also~\cite{Briaudeau1998}).

\begin{enumerate}

\item The first term $J_1$ is the one we would have obtained had we simply taken for
the susceptibility $\chi(z,z',\omega,v)$ of atoms at velocity $v$ the local, Doppler-shifted (and incorrect!) expression:
\beq\label{Eq:chi_local}
\chi(z-z',\omega,v)= i{NC_{FF'}^2d^2\over \hbar\epsilon_0}{M_v(v)\over  \Gamma_{\rm t}/2-i(\Delta_{FF'}-k_{\rm l}v)}\, \delta(z-z')
\eeq
where the detuning $\Delta_{FF'}=\omega- \omega_{FF'}$. Although intuitive, this approach would have missed
the contributions from the two remaining terms, which are important in the mesoscopic regime.
The term $J_1$ dominates as soon as $L\gtrsim |v/\Lambda|$. Close to resonance and for $kv\lesssim \Gamma_{\rm t}$,
this yields $L\gtrsim \xi$.
This indicates that in large cells ($L\gtrsim 1\,{\mu}$m), $J_1$ is the dominant contribution and the conventional dispersion theory applies.

\item The second and third terms $J_2$ and $J_3$ become important as soon as $|v|/(|\Lambda|L)\gtrsim 1$, i.e. $L\lesssim \xi$.
These two terms are specicific to the mesoscopic regime and are a consequence of the non-local  character
of the atomic response. They dominate for thin cells.

\item Finally, the third term $J_3\sim J_2 \exp[-L/ \xi]$ describes the fact that the atomic dipole stops
emitting when it reaches the boundary of the slab. When $J_2\gtrsim J_1$,
we have seen that $L<\xi$ and $J_3\sim J_2$,
indicating that in the mesoscopic regime it is not possible that $J_2$ alone dominates.

\end{enumerate}

\subsection{Transmitted field in a dilute regime with cavity boundaries}\label{SubSec:slab_sapphire}

\begin{center}
\begin{figure}
\includegraphics[width=0.8\columnwidth]{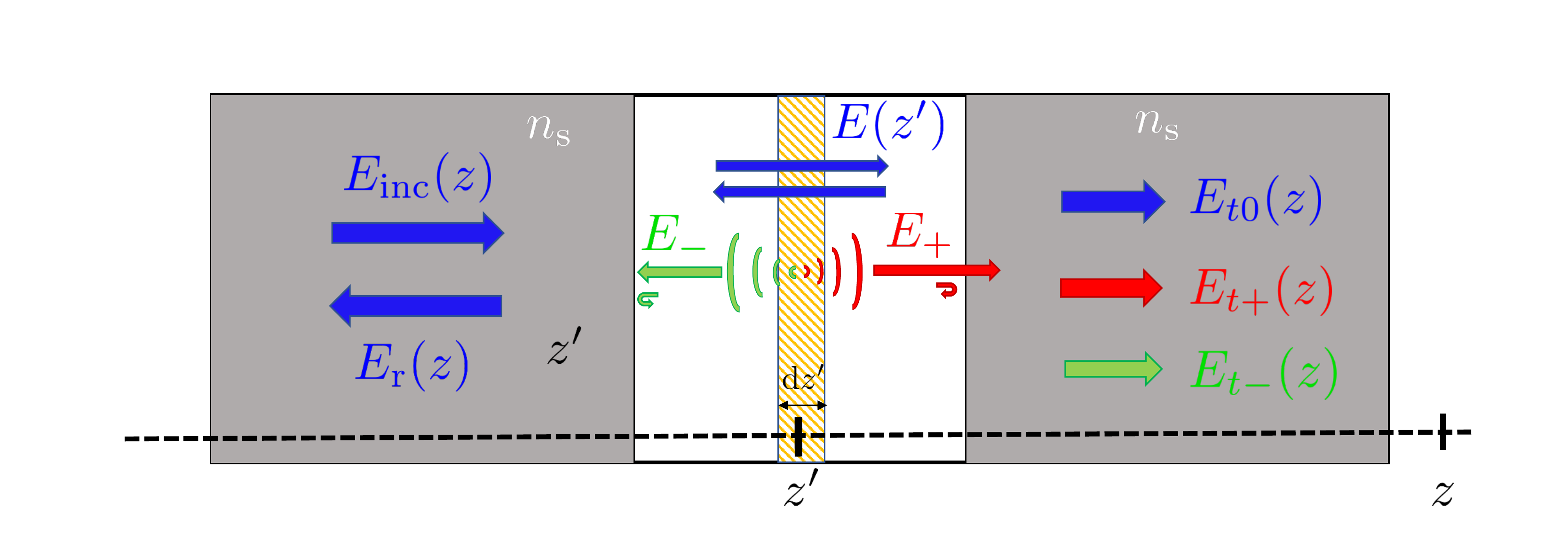}
\caption{Configuration used to calculate the transmission through the cell.
The dilute vapour is confined between two sapphire plate of index $n_{\rm s}$.
The incident field $E_0\exp[ik_{\rm l}z]$ undergoes multiple reflections inside the cell before
being finally transmitted ($E_{t0}$).
The field $E(z')$ inside the cell excite the atoms in the slice $dz'$, and radiate
a field in the forward (backward) direction $E_+$ ($E_-$)
that is reflected multiple times before leading to the transmitted fields
$E_{t+}$ ($E_{t-}$).}
\label{FigSM0}
\end{figure}
\end{center}

We now consider the more involved case where the atomic slab is confined between the two windows of the cell
(index $n_{\rm s}$), leading to a pronounced cavity effect. This situation has already been discussed,
in particular in Ref.~\cite{Dutier2003a} starting from a different point of view. Here we use the non-local susceptibility
given by Eq.\eqref{chi2} to derive the transmission of the slab confined in the nanocell.

Figure~\ref{FigSM0} summarizes the different fields involved.
First, we need to consider the field inside the empty cavity between the two windows, which reads:
\beq\label{cavityfield}
E(z'')=\frac{E_0t_1}{1-r_2^2e^{2ik_{\rm l}L}}\left( e^{ik_{\rm l}z''}+r_2e^{ik_{\rm l}(2L-z'')} \right)
\eeq
with $t_1=2n_{\rm s}/(1+n_{\rm s})$ and $r_2=(1-n_{\rm s})/(1+n_{\rm s})$.
Second, the field emitted by the atoms of the vapor also undergoes  multiple reflections
on the cavity walls. We separate this field into two components propagating respectively
in the forward and backward directions and consider the multiple reflections of each.
Consequently, the field transmitted after the cell results from  the interference of
(i) the field $E_{\rm t0}= \frac{t_1t_2}{1-r_2^2e^{2ik_{\rm l}L}} E_{0}e^{ik_{\rm l}z}$ of the cavity without any atoms,
(ii) the field $E_{\rm t+}$ emitted by all slices $dz'$ initially directed in the forward direction and undergoing multiple reflections, and
(iii) the field $E_{\rm t-}$ initially emitted in the backward direction, also undergoing reflections.
The transmitted field is therefore:
\beq \label{totalfield1}
E_{\rm t}(z)=\frac{t_1t_2}{1-r_2^2e^{2ik_{\rm l}L}} E_{0}e^{ik_{\rm l}z}+E_{\rm t+}(z)+E_{\rm t-}(z)\ ,
\eeq
with $t_2=2/(n_{\rm s}+1)$. The field emitted by the atoms in the forward direction is,
taking into account the non-local relation between $P(z)$ and $E(z)$ inside the vapor:
\beq\label{Eqforward}
E_{\rm t+}=A\frac{ik_{\rm l}}{2\epsilon_0}\int_{0}^{L}dz'e^{ik_{\rm l}(z-z')}\left[ \int_{-\infty}^{\infty}dz''\epsilon_0 \left(\int_{-\infty}^{\infty}dv \,\chi_{_L}(z',z'',\omega,v)\, \right)E(z'')\right]
\eeq
with
\beq\label{FPinside}
A=t_2 (1+r_2^2e^{2ik_{\rm l}L}+r_2^4e^{4ik_{\rm l}L}+...)=\frac{t_2}{1-r_2^2e^{2ik_{\rm l}L}}
\eeq
the prefactor that accounts for the multiple reflections before the field exits the cavity.
To calculate the integral we again assume a dilute vapor and use the expression \eqref{cavityfield}
valid for an empty cavity. Calculations similar to the ones performed for the slab immersed in vacuum
(Sec.\ref{Sec:slab_vacuum}) lead to:
\beq\label{eqEtf}
E_{\rm t+}=A B C \left(I_1+I_2\right)E_0\exp[ik_{\rm l}z]
\eeq
with
\beq\label{Eq:I1}
I_1=-\int_{0}^{\infty}dv M_v(v) \bigg[\frac{L}{\Lambda_{+}}+
\frac{v}{\Lambda_{+}^2}e^{-\Lambda_{+}{L\over v}}-\frac{v}{\Lambda_{+}^2}
+\frac{r_2e^{2ik_{\rm l}L}}{\Lambda_{-}}\left( \frac{1}{2ik_{\rm l}}(1-e^{-2ik_{\rm l}L})+
\frac{v}{\Lambda_{+}}(e^{-\Lambda_{+}{L\over v}}-1)\right)\bigg]\ ,
\eeq
and
\beq\label{Eq:I2}
I_2=\int_{-\infty}^{0}dv M_v(v)\bigg[-\frac{L}{\Lambda_{+}}+\frac{v}{\Lambda_+^2}e^{\Lambda_+{L\over v}}
-\frac{v}{\Lambda_+^2}\\
+\frac{r_2e^{2ik_lL}}{\Lambda_-}\left(\frac{1}{2ik_{\rm l}}(e^{-2ik_{\rm l}L}-1)+\frac{v}{\Lambda_+}
e^{\Lambda_-{L\over v}}(1-e^{-\Lambda_+{L\over v}})\right) \bigg]\ ,
\eeq
where we have introduced $\Lambda_{\pm}=\Gamma_{\rm t}/2-i(\Delta_{FF'} \mp k_{\rm l}v)$, $B=t_1/(1-r_2^2e^{2ik_{\rm l} L})$ and
$C=NC_{FF'}^2d^2k_{\rm l}/(2\hbar\epsilon_0)$.

We proceed in a similar way for the field $E_{\rm t-}$ initially emitted in the backward direction, noting that
the polarization vector at position $z'$ due to the backward emitted field is retarded by a factor
$r_2e^{2ik_{\rm l}z'}$ (see Fig.~\ref{FigSM0}). We get:
\beq\label{Eqbackward}
E_{\rm t-}=A r_2\frac{ik_{\rm l}}{2\epsilon_0}\int_{0}^{L}dz'e^{ik_{\rm l}(z+z')}\left[ \int_{-\infty}^{\infty}dz''\epsilon_0
\left(\int_{-\infty}^{\infty}dv \,\chi_{_L}(z',z'',\omega,v) \right)E(z'')\right]\ .
\eeq
This leads to:
\beq\label{eqEtb}
E_{\rm t-}=r_2 A B C(I_3+I_4)E_0\exp[ik_{\rm l}z]
\eeq
with
\beq\label{Eq:I3}
I_3=-\int_{0}^{\infty}dv M_v(v)
\bigg[\frac{1}{\Lambda_{+}}\left({e^{2ik_{\rm l}L}-1\over 2ik_{\rm l}}+\frac{v}{\Lambda_{-}}(e^{-\Lambda_{-}{L\over v}}-1)\right)
+\frac{r_2e^{2ik_{\rm l}L}}{\Lambda_{-}}\left( L+\frac{v}{\Lambda_{-}}(e^{-\Lambda_{-}{L\over v}}-1)\right)\bigg]
\eeq
and
\beq\label{Eq:I4}
I_4=\int_{-\infty}^{0}dv M_v(v) \bigg[\frac{1}{\Lambda_{+}}\left({1-e^{2ik_{\rm l}L}\over 2ik_{\rm l}}+\frac{v}{\Lambda_{-}}(e^{\Lambda_{+}{L\over v}}-e^{2ik_{\rm l}L}))\right)
+\frac{r_2e^{2ik_{\rm l}L}}{\Lambda_-}\left(-L +\frac{v}{\Lambda_{-}}(e^{\Lambda_{-}{L\over v}}-1)\right)\bigg]\ .
\eeq
Finally, the transmitted field is:
\beq\label{Equation_final}
E_{\rm t}(z)=\left(\frac{t_1t_2}{1-r_2^2e^{2ik_{\rm l}L}}+ ABC\left[I_1+I_2 +r_2(I_3+I_4)\right]\right)E_0\exp[ik_{\rm l}z]\ .
\eeq
We retrieve the  expression obtained in Ref.~\cite{Dutier2003a} using a different approach.
Here again, the fact that in Eqs.\,\eqref{Eqforward} and \eqref{Eqbackward} we can separate the integral
into two integrals involving only the positive [Eqs.\eqref{Eq:I1} and \eqref{Eq:I3}] or the negative
velocities [Eqs.\eqref{Eq:I2} and \eqref{Eq:I4}] comes from the assumptions
that the atomic coherences are lost in a quenching collisions at the cell walls.

\subsection{Case of a multilevel atom}

So far we have considered a single atomic transition. In the experiment, several hyperfine transitions of the
D1 cesium line contribute. In order to account for them we calculate the susceptibility by summing
over all transitions and  Eqs.\,\eqref{transFin1} and
\eqref{Equation_final} respectively become:
\beq\label{multi1}
E_{\rm t}=E_0e^{ik_{\rm }lz} \left[1-\sum_{FF'}\frac{NC_{FF'}^2d^2}{4\hbar\epsilon_0 }\int_{-\infty}^{\infty} dv M_v(v) \left( \frac{k_{\rm l}L}{\Lambda}-\frac{k_{\rm l}|v|}{\Lambda^2}+\frac{k_{\rm l}|v| e^{-{\Lambda L \over |v|}}}{\Lambda^2}\right) \right]\ ,
\eeq
\beq\label{multi2}
E_{\rm t}(z)=\left(\frac{t_1t_2}{1-r_2^2e^{2ik_{\rm l}L}}+\sum_{FF'}ABC\left[I_1+I_2 +r_2(I_3+I_4)\right]\right)E_0\exp[ik_{\rm l}z]\ ,
\eeq
where the detuning appearing in $\Lambda$ is $\Delta_{FF'} = \omega-\omega_{FF'}$.

\section{Bimodal velocity distribution}\label{Sec:bimodal}

At the end of the main text, we compare two models: a phenomenological model where we use the local susceptibility, but modify the velocity distribution to account for the non locality due to the interface, taking $M_v(v)=\delta(v)$, and the more accurate non-local model derived in Sec.\ref{Sec:NLmodel}.
We conclude that the second non-local model is able to explain the data in a more satisfactory way, as it yields a linewidth $\Gamma_{\rm p}$ independent of $L$.
We show here that the use of a more refined velocity distribution in  phenomenological model does not affect this conclusion.

\begin{figure}
\includegraphics[width=0.5\columnwidth]{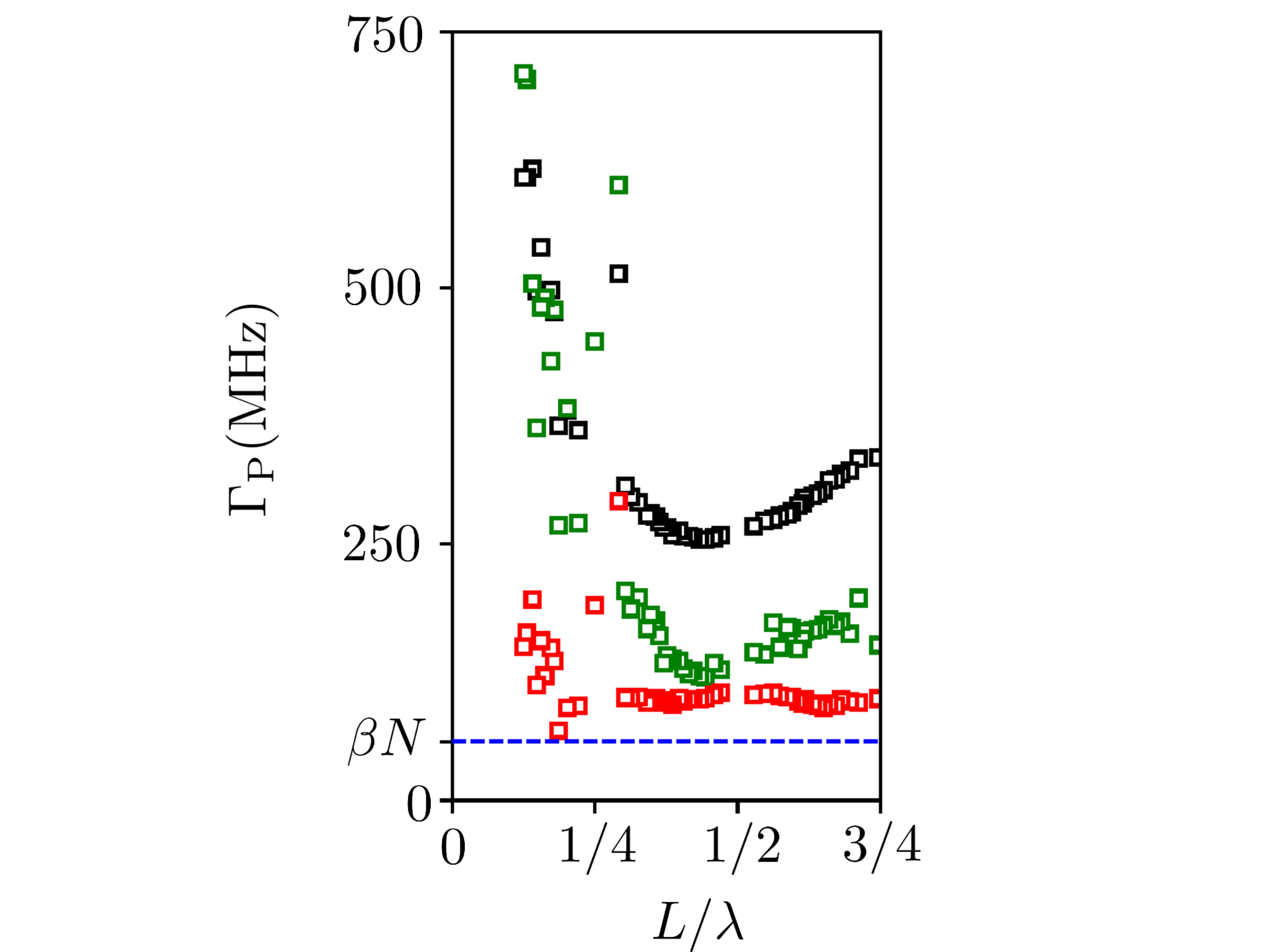}
\caption{Fitted broadening parameter $\Gamma_{\rm p}$ versus the cell thickness $L$.
 Black squares: phenomenological model where we use the local susceptibility
with $M_v(v)=\delta(v)$.
Green squares: same  using the bimodal velocity distribution.
Red squares: accurate non-local model  as derived in Sec.~\ref{Sec:NLmodel}. }
\label{FigSM3}
\end{figure}

To do so, we use a bimodal normalized velocity distribution, already introduced in the context of nano-cells~\cite{Whittaker2015}:
\begin{equation} \label{Bimodal}
M_v(v)=W\left(a\exp\left[-{v^2\over u^2}\right]+(1-a)\exp\left[-\left({v\over su}\right)^2\right]\right)\ ,
\end{equation}
where $u=\sqrt{2k_{\rm B}T/M}$ is the root-mean square velocity and $W=\left( au\sqrt{\pi}+(1-a)su\sqrt{\pi}\right)^{-1}$.
The additional parameters $a$ and $s$ describe respectively the fraction of slow atoms and the width of the component of
the distribution corresponding to the fast atoms.
This bimodal distribution therefore consists in  a  pedestal associated to the slow atoms
and a sharp feature superimposed on coming from the fast atoms.
We fit the data using this velocity distribution in the local model with
$N$, $\Gamma_{\rm p}$, $\Delta_{\rm p}$, $a$ and $s$ as free parameters. The result is shown in Fig.~\ref{FigSM3}
as green squares. Similarly to the case where $M_v(v)=\delta(v)$, the strong dependence of  $\Gamma_{\rm p}$ with $L$
is not compatible with a physical interpretation.

\section{Deviation from Beer-Lambert transmission law}\label{Sec:Beer}

\begin{figure}
\includegraphics[width=0.5\columnwidth]{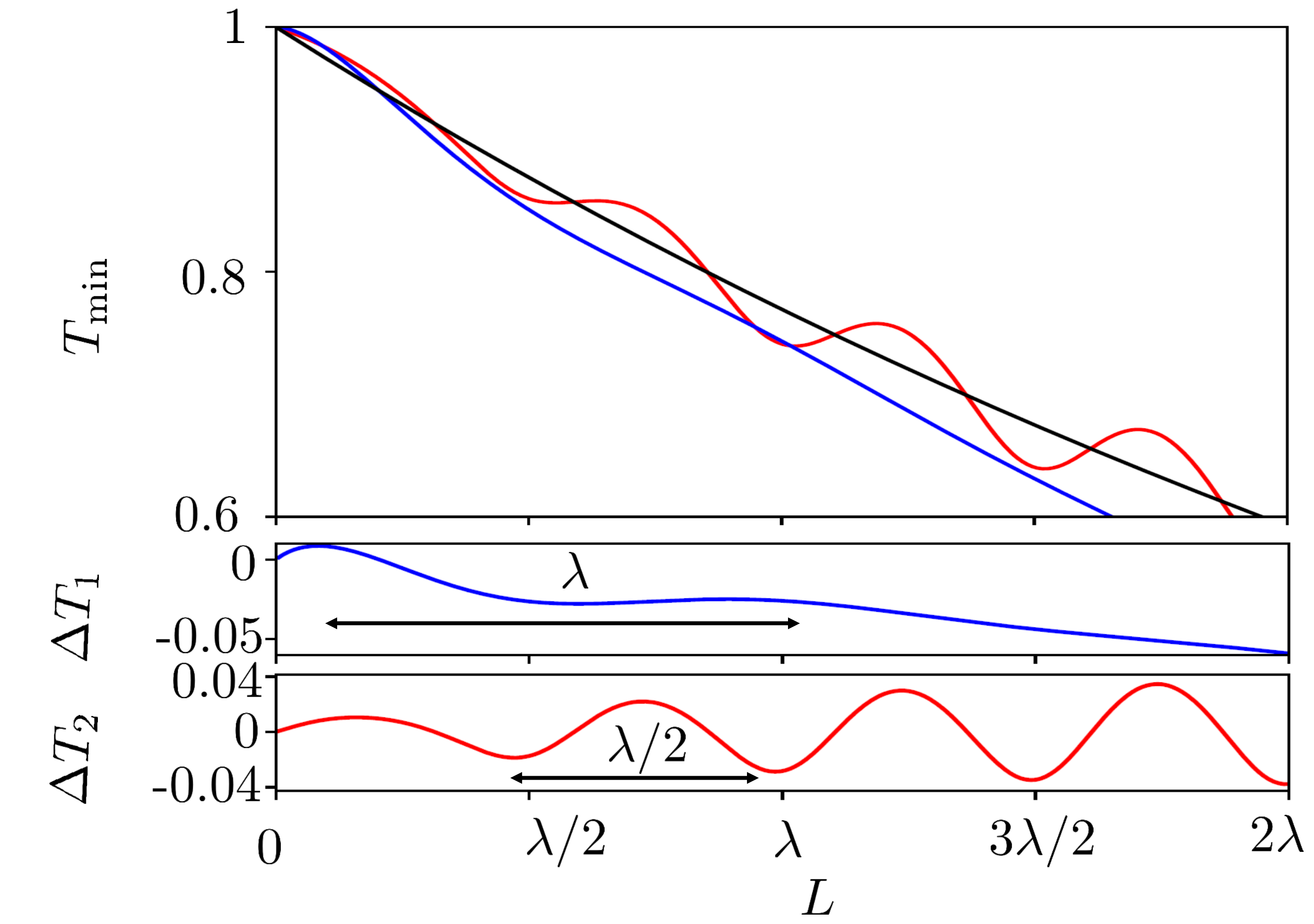}
\caption{Top : Minimum of the transmission, $T_{\rm min}$, versus $L$ extracted from the
local model including the cavity (red line) and from the non-local model without any cavity
surrounding the slab (blue line). Black line : prediction of the Beer-Lambert law.
Parameters used : $\Delta_{\rm p}=0$, $\Gamma_{\rm p}=2\pi\times 100$\,MHz and $\Theta=188^{\circ}$C.
Bottom : deviations of the non-local ($\Delta T_1$) and local ($\Delta T_2$)
models from the Beer Lambert law.}
\label{FigSM4}
\end{figure}

Figures 3(a,b) of the main text show that the minimum of transmission of the hyperfine  transition $F=4$ to $F'=3$
does not follow the Beer-Lambert law. To gain further insight on this behavior, we study the
evolution of the minimum of the theoretical transmission $T_{\rm min}$ as a function of the vapor thickness $L$.
To separate the effect of the cavity from the one of the non-local response, we consider two models:
(i) the first local model of the text that accounts for the cavity (Eq.\,(3) of main text) with a local index of refraction.
(ii) the non-local model, but without the cavity surrounding the atomic (Eq.\,\eqref{totalfield1} of the Supplemental Material).
The results are shown in Fig.~\ref{FigSM4}.
We observe  that the cavity surrounding a local-medium leads to an oscillation of
$T_{\rm min}$ with $L$ with a period $\lambda/2$,
as it should for a standing wave in a cavity.
The non-local response of the cloud results into an oscillation of $T_{\rm min}$ with a period $\lambda$.
This can also be seen from Eq.\,\eqref{transFin1} where the last term $\exp[-\Lambda L/|v|]$
leads to an oscillatory dependence in $k_{\rm l}L$.

\section{Thickness dependence of $\Gamma_{\rm p}$ in the phenomenological second model}

\begin{figure}
\includegraphics[width=0.5\columnwidth]{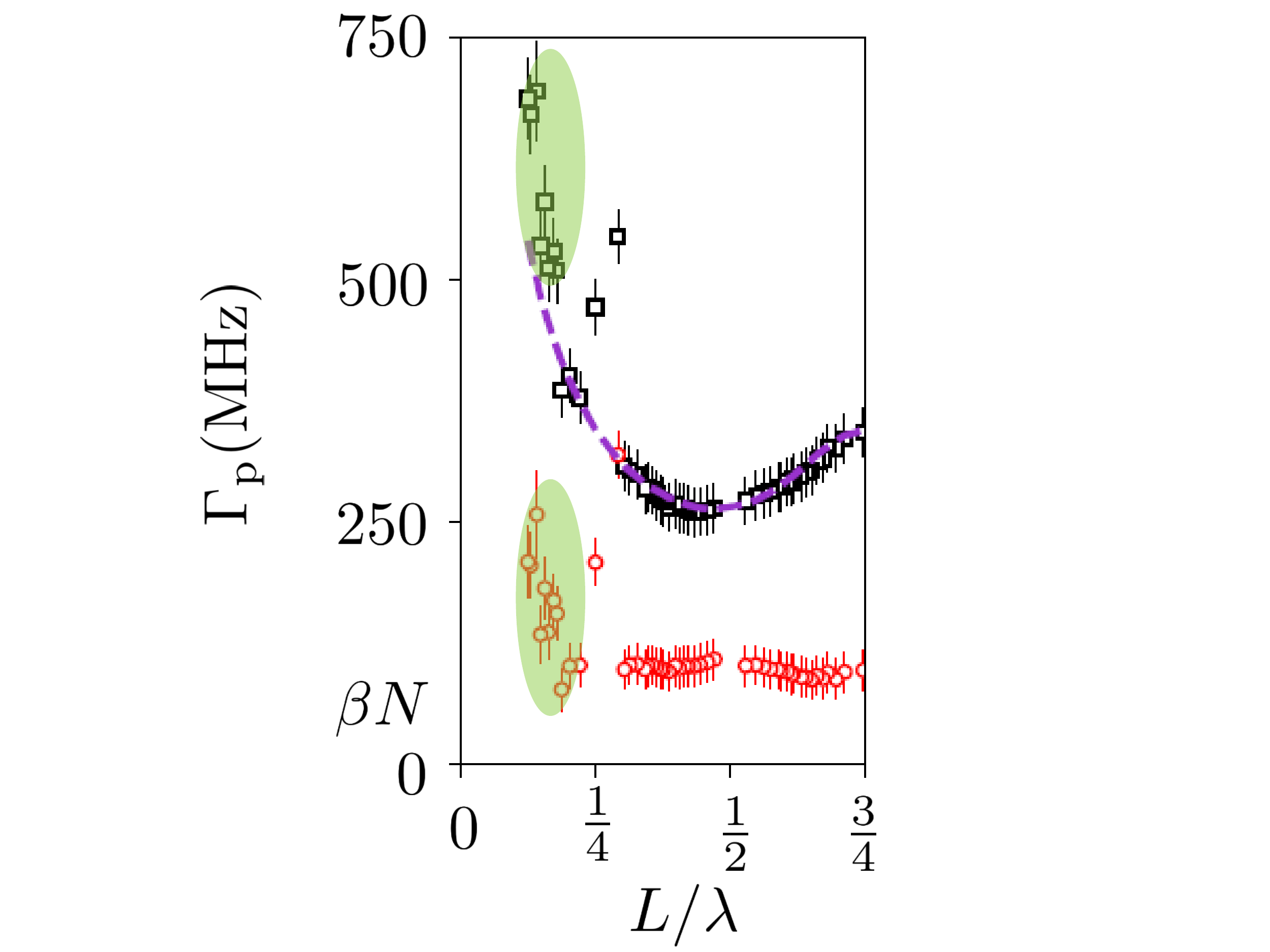}
\caption{Broadening parameter $\Gamma_{\rm p}$  obtained from the fit of the experimental spectra using the  {\it phenomenological} second model  of the main text (black  squares)
and the  non-local model of Sec.~\ref{Sec:NLmodel} (red  circles). Error bars are the quadratic sum of statistic and fit errors. The violet dashed line is the broadening parameter $\Gamma_{\rm p}$  obtained from the fit using the phenomenological non-local model of the data simulated by the expression derived in Sec.~\ref{SubSec:slab_sapphire}.
Green shaded areas indicate the region where the Casimir atom-surface interactions dominate the broadening.}
\label{FigSM5}
\end{figure}

In Fig.3(c) of the main text, we observe a strong dependence of
the broadening parameter $\Gamma_{\rm p}$ with the thickness $L$ for the phenomenological model. In this section we deconvolve the effects of the Casimir atom-surface  interaction from the non-local influence of the surface when $\xi \geq L$ on the parameter $\Gamma_{\rm P}$ returned by the fit.
To understand the sharp increase for small $L$, we simulate  a collection of spectra with the accurate non-local model of Sec.~\ref{Sec:NLmodel}
for a fixed temperature
($\Theta=170^{\circ}$C), a fixed broadening
$\Gamma_{\rm p}=100$\,MHz and various $L$ between $\lambda/8$ and $3\lambda /4$.
We do not include the Casimir atom-surface interaction in the model.
We then fit  each generated spectra by the {\it phenomenological} model and extract the corresponding
$\Gamma_{\rm p}$. The result is shown as a violet dashed line in Fig.~\ref{FigSM5}.

 We observe a very good agreement of the fitted $\Gamma_{\rm p}$ for
the experimental and simulated data using the accurate non-local model
for $L\gtrsim 150$\,nm. This indicates that the increase of $\Gamma_{\rm p}$ at small $L$
is a feature of the phenomenological, non-local model. The
sharper increase for $L\leq150$\,nm is therefore ascribed to the Casimir atom-surface interactions,
and can also be seen in the variation of $\Gamma_{\rm p}$ returned by the accurate non-local model with $L$.

